\newcommand{\tsecompldate}{22nd November 1999}

\documentclass[12pt,a4paper]{article}

\usepackage{amsfonts}

\usepackage{graphics}
\newcommand{\journal}[1]{#1}
\newcommand{\vol}[1]{{\bf #1}}

\newcommand{\ttitle}[1]{{\it #1}}
\newcommand{\tpretitle}[1]{}
\newcommand{\arttitle}[1]{}
\newcommand{\inproctitle}[1]{``#1''}

\newcommand{\tref}[1]{(\ref{#1})}

\newcommand{\tnote}[1]{}
\newcommand{\tcomment}[1]{}
\newcommand{\tnotpre}[1]{#1}
\newcommand{\tpre}[1]{}
\newcommand{\tprenote}[1]{}

\newcommand{\href}[2]{#2}
\newcommand{\eprint}[1]{{\tt #1}}

\newcommand{\tsedevelop}[1]{{}}

\renewcommand{\tnotpre}[1]{}
\renewcommand{\tpre}[1]{#1}
\renewcommand{\tprenote}[1]{\footnote{#1}}
\renewcommand{\href}[2]{{#2}{}}
\renewcommand{\eprint}[1]{\href{http://xxx.soton.ac.uk/abs/#1}{{\tt #1}}}
\renewcommand{\vol}[1]{{\bf #1}}

\renewcommand{\journal}[1]{#1}
\renewcommand{\ttitle}[1]{{\em #1}}
\renewcommand{\tpretitle}[1]{{\em #1}}

\renewcommand{\inproctitle}[1]{``#1''}
\renewcommand{\arttitle}[1]{{\em #1},}

\newcommand{\half}{\frac{1}{2}}
\newcommand{\bea}{\begin{eqnarray}}
\newcommand{\eea}{\end{eqnarray}}
\newcommand{\beq}{\begin{equation}}
\newcommand{\eeq}{\end{equation}}

\newcommand{\jvec}{\vec{j}}
\newcommand{\phivec}{\vec{\phi}}

\newcommand {\expectn}[1]{\ensuremath{\big\langle #1
\big\rangle}}

\newcommand {\ao}[1]{\langle {#1} \rangle_{0}}

\newcommand {\av}[1]{{\langle {#1} \rangle}}

\begin{document}

\renewcommand{\thefootnote}{\fnsymbol{footnote}}

\tpre{\begin{flushright} {\tt Imperial/TP/99-0/09} \\
\eprint{hep-ph/9911492} \\ Present version \tsecompldate \\
\tsedevelop{ (LaTeX-ed on \today ) \\}
%{next preprint number}\\
\end{flushright}
\vspace*{1cm}}

\begin{center}
{\Large\bf An Optimised Perturbation Expansion} \\ {\Large\bf  for
a Global O(2) Theory}\\ \tpre{\vspace*{1cm} } {\large
T.S.Evans\footnote{email: \href{mailto:T.Evans@ic.ac.uk}{{\tt
T.Evans@ic.ac.uk}}\tpre{, WWW:
\href{http://euclid.tp.ph.ic.ac.uk/links/time}{\tt
http://euclid.tp.ph.ic.ac.uk/\symbol{126}time} }},
M.Ivin\footnote{email: \href{mailto:M.Ivin@ic.ac.uk}{{\tt
M.Ivin@ic.ac.uk}}}, M.M\"{o}bius\footnote{email:
\href{mailto:memobius@midway.uchicago.edu}{{\tt
memobius@midway.uchicago.edu}}, present address: Enrico Fermi
Institute, University of Chicago}}
\\
\tpre{\vspace*{1cm}}
\href{http://euclid.tp.ph.ic.ac.uk/}{Theoretical Physics},
Blackett Laboratory, Imperial College,
\\
Prince Consort Road, London, SW7 2BZ,  U.K.
\tnotpre{\\ Tel: U.K.-20-7594-7837,
Fax: U.K.-20-7594-7844 (or -7777)
\\
PACS: 11.10.Gh,  02.30.Cj,  02.30.Tb
\\
Key Words: delta expansion, symmetry breaking, non-perturbative
methods}
\end{center}

\begin{abstract}
We use an optimised perturbation expansion called the linear
$\delta$-expansion to study the phase transition in a Higgs sector
with a continuous symmetry and large couplings.  Our results show
how to use this non-perturbative method successfully for such
problems.  We also show how to simplify the method without losing
any flexibility.
\end{abstract}

%02.30.Cj Measure and integration
%02.30.Tb Operator theory
%11.10.-z  Field theory (for gauge field theories, see 11.15)
%11.10.Gh Renormalization

{}\tnote{tnotes such as this not present in final version}

\renewcommand{\thefootnote}{\arabic{footnote}}
\setcounter{footnote}{0}

% *******************************************************
% *******************************************************

Quantum Field Theory has had many successes over the last fifty
years but most are based on the use of small coupling perturbation
expansions.  However, there are many interesting problems where
such an expansion is not appropriate.  QCD is the classic example
but phase transitions at non-zero temperatures even in models with
small coupling constants, e.g.\ the Electro Weak model, are now
known to be in this class. It is this latter example which
motivates our work which will focus on the non-perturbative
behaviour of the Higgs sector.

Alternatives to perturbation theory, non-perturbative methods,
have many limitations.  For instance it is usually very difficult
to go beyond the lowest order in methods such as large N
\cite{Raybook} while numerical Monte Carlo codes are relatively
expensive \cite{Mo,Lu}.\tnote{They are also unable to handle
systems with a non-zero charge, see Montevay and M\"unster,
Lattice Gauge Theory book.} Another method with a long history is
the Linked Cluster Expansion \cite{Wo,LW} which now also exploits
large amounts of computing power \cite{HR}. In this letter we
extend a non-perturbative method called an OPE (Optimised
Perturbation Expansion).  In this approach one expands around a
solvable model which contains several arbitrary parameters, and
thus is a variational ansatz.  The expansion is not necessarily in
terms of some small parameter.  One then chooses the variational
parameters, invariably using some optimisation criteria which is a
highly non-linear procedure. The optimisation criteria has to
ensure good results from a few terms of the series but there has
been much discussion about what is a good optimisation criteria.

As described, the OPE is a very general method so not surprisingly
it has been rediscovered several times in different guises,
applied to very different applications, and appears under many
different names; it is also known as the linear $\delta$-expansion
\cite{Jo}, action-variational approach \cite{KM}, improved
gaussian approximation \cite{Ok}, variational perturbation theory
\cite{SSS}, method of self-similar approximation \cite{Yu}, or the
variational cumulant expansion \cite{WZZSDYX}.  For instance the
method has been to the evaluation of simple integrals
\cite{BuDJ,BeDJ,Jo}, solving non-linear differential equations
\cite{BMPS}, quantum mechanics \cite{Ca,Ki,DJ,Ok,Jo} to quantum
field theory both in the continuum \cite{Ok,SSS} and on a lattice
\cite{Jo,AJ93a,AJ93b,AJP,ZTW,Ya,YWZ,ZL,EJR98a,EJR98b,KM,WZZSDYX}.

To motivate this work we first note that the analysis of pure
gauge theories on the lattice using OPE used a variety of
optimisation schemes when fixing the variational parameters.  One
can minimise the free energy \cite{ZTW} or demand that the series
converges as fast as possible by minimising the high order terms
with respect to the lower order terms \cite{KM} - the principle of
fastest convergence.  However, we believe the best results for
pure gauge models are obtained when using the principle of minimal
sensitivity \cite{St}, as discussed in \cite{Jo,AJ93a,AJ93b,AJP}.
In this case the variational parameters, say $\{ \alpha \}$, are
set {\em separately} for each physical quantity, say $O$, under
consideration by demanding that the quantity changes as little as
possible if the variational parameters $\{ \alpha \}$ are varied
from their optimal values $\{ \tilde{\alpha} \}$, specifically
\begin{equation}
\left.
\frac{\partial \langle O \rangle}{\partial \alpha_i}
\right|_{\alpha=\tilde{\alpha}}= 0 .
\label{PMS}
\end{equation}
Combinations of these criteria are also possible \cite{Yu}.

Given the interest in non-perturbative effects in gauged Higgs
models it is interesting to note that the situation for the OPE in
these cases \cite{Ya,YWZ,ZL,EJR98a,EJR98b} is not nearly so good,
and the results are not as extensive as for the pure gauge case.
The analysis of the gauged U(1) Higgs model \cite{Ya,YWZ} fixed
the variational parameters by minimising the {\em lowest} order
calculation of the free energy, a procedure which is justified by
noting the identity\tnote{Kleinert calls this the Jensen-Peierls
identity.} $\langle {\exp \{ -O \} } \rangle \geq \exp \{ - \langle
{O} \rangle  \}$. However, this approach was rejected in the pure
gauge analyses and we will not pursue it here.  For the more
interesting gauged SU(2) Higgs model the analysis in
\cite{EJR98a,EJR98b} was quite successful in reproducing the known
Monte Carlo data but the accuracy obtained fell short of that
obtained when using similar methods for pure gauge
models.\tnote{We put the results of \cite{ZL} to one side in this
discussion. The authors of \cite{EJR98a,EJR98b} were able to
reproduce the results given in \cite{ZL} but only by using a very
strange criterion to fix the variational parameters.} This is
understandable as the analysis revealed that adding the Higgs
sector produces new problems (as indeed it does for numerical lattice Monte
Carlo studies), and the variational ansatz used was relatively
simple.

Given that the Higgs sector causes new problems, it is worth
considering pure Higgs models on the lattice in the context of OPE
methods, effectively the zero gauge coupling limit of the gauged
Higgs models rather than the infinite Higgs mass limit represented
by the pure gauge models.  However, we have found only a
discussion of the {\em real} scalar model on the lattice using OPE
method \cite{WZZSDYX} rather than models with continuous global
symmetries relevant to realistic gauged Higgs models.

The purpose of this paper is to present results for the simplest
prototype of a Higgs sector in a gauge theory, a pair of scalar
fields with an O(2) global symmetry.  It is hoped that the lessons
learned in this case will improve future work on realistic
problems.  The model has not been studied before using the OPE on
the lattice and further the Higgs sector in this model is
different from that used in \cite{EJR98a,EJR98b} where only the
modulus of the Higgs was required. Likewise the variational ansatz
we choose here is more complicated with more variational
parameters than that used in \cite{WZZSDYX} or
\cite{EJR98a,EJR98b}.  We also study different schemes to
fix the variational parameters.

Our results show that the OPE can work well for problems with
global continuous symmetry and large coupling constants. We
highlight the role of the symmetry in the choice of the
variational ansatz and how to use this to reduce the number of
necessary variational parameters. We also show that minimising the
free energy alone is sufficient to locate the phase transition.
Further using the minimum of the free energy to fix the
variational parameters for calculations of {\em any} quantity
produced sensible results. This is to be set against our results
that show we were unable to find turning points of the form
\tref{PMS} for several different types of expectation value.  Thus
it appears that the principle of minimal sensitivity \cite{St}
does not work in this model in contrast with its great success in
pure gauge models \cite{Jo,AJ93a,AJ93b,AJP} and the rigorous
results obtained with this approach to OPE in other contexts
\cite{BuDJ,BeDJ,DJ}.

Our starting point is the Euclidean action in four-dimensions for
a scalar field theory with a global O(2) symmetry, or equivalently
a global U(1) symmetry, put on a four-dimensional hyper-cubic space-time lattice to
regulate the UV divergences
\begin{equation}\label{action}
S(\{\phi_{i},\psi_{i} \})
=
- \sum_{n}^{N}\sum_{\mu}^{4} (\phi_{n+\mu}\phi_{n}+\psi_{n+\mu}\psi_{n})
+ \sum_{n}^{N}m^{2} (\phi_{n}^{2}+\psi_{n}^{2})
+ \sum_{n}^{N}\lambda (\phi_{n}^{2}+\psi_{n}^{2})^{2} .
\end{equation}
The subscript $n$ runs over all $N$ lattice sites and the $n+\mu$
subscript stands for the nearest neighbour terms.
The aim is to calculate the partition function or generating functional, $Z$,
\begin{equation}\label{Zdef}
Z=\prod_{i=1}^{N}\int \textrm{d}\phi_{i} \textrm{d}\psi_{i}
e^{-S(\{\phi_{i},\psi_{i} \})}
\end{equation}
for large couplings $\lambda$.  We will also be interested in
various expectation values such as $\expectn {\phi}$ and $\expectn
{\phi^2}$.

The OPE method we will use to perform such calculations
is the linear $\delta$-expansion \cite{Jo}. A suitable trial action
$S_{0}$ is added and subtracted from the real action, so that we
use the action
$S_\delta = S_{0} + \delta(S- S_{0})$ and $\delta=1$, i.e.\
\begin{eqnarray}\label{delta}
Z
&=&
\sum_{n=0}^{\infty} Z_0 \frac{\delta^n}{n!} \ao{(S_{0}-S)^n}
\end{eqnarray}
where $\ao{\ldots}$ is called the statistical average defined as
\begin{equation}\label{statav}
\ao{\ldots} := \frac{1}{Z_{0}}\prod_{i=1}^{N}\int \textrm{d}
\phi_{i} \textrm{d} \psi_{i} (\ldots) \exp \{-S_{0}\} .
\end{equation}
$Z_{0}$ is the partition function for the action $S_{0}$. Note
that $\delta$ is just a formal counting parameter since it must be
set to one to make contact with the physical theory. It is used to
help us truncate the infinite number of Feynman diagrams which
represent the solutions to QFT, i.e.\ we will work up to a finite
power of $\delta$.  It seems that the ultimate success of any
calculation will rely on the small size of $\ao{(S_{0}-S)^n}$.
Thus we might guess that, as with the ansatz in any variational
approach, $S_0$ needs to be as good a representation of the
correct physics as possible.  However, this requires us to know
the full answer before we can make a good guess for $S_0$.  To
break this circle, we use our physical intuition to make the best
guess for  the general {\em form} of $S_0$ but at the same time we
let $S_0$ depend on some unphysical variational parameters. When
we truncate our expressions such as \tref{delta} at some finite
order in delta, our answers will depend on these parameters and we
will optimize our values for physical quantities by choosing
values for these unphysical parameters using suitable criteria.
Thus we are optimising our choice of $S_0$ using a variational
method and it is the optimization which introduces the highly
non-linear and non-perturbative element of OPE.

In our case we choose our trial action, $S_0 = S_0(\phi,\psi ;
m^2, \lambda ; j_\phi,j_\psi , k_+, k_-)$, to be
\begin{equation} \label{TrialS}
S_0 =
- \sum_{n} \! \left[
      j_{\phi} \phi_n + j_{\psi} \psi_n
   +  \half k_+ ( \phi_n^2 + \psi_n^2)
   +  \half k_- ( \phi_n^2 - \psi_n^2)
   -  \lambda (\phi_n^2 + \psi_n^2)^2
\right]
\end{equation}
introducing the unphysical variational parameters
$j_\phi,j_\psi,k_+,k_-$.  We do not introduce a $k \phi\psi$ term
as we can always make an O(2) rotation of our fields to eliminate
such a term.  We therefore assume that this has been done so that
in general we have used up our freedom to make such a field
redefinition. Our $S_0$ is then a reasonable guess for an action
with two scalar degrees of freedom since the form of $S_0$ is very
similar to $S$ and their ultra local parts are identical when
$j_\phi=j_\psi=k_-=0$ and $k_+=-m^2$.  We therefore might hope
that $\ao{(S_{0}-S)^n}$ is likely to be small, and indeed perhaps
minimised for values of the variational parameters where the
ultralocal parts of $S$ and $S_0$ are equal.   Note that our $S_0$
is significantly different from that of \cite{EJR98a,EJR98b} as
when $k_- \neq 0$ or either $j\neq 0$ then this $S_0$ is not O(2)
symmetric.   We shall also work with the full complex field (or
its equivalent) rather than just its modulus as used in the gauged
model in \cite{EJR98a,EJR98b}.

However, there is a second criterion for the choice of $S_0$,
namely it must be of a form which allows us to perform some
calculations.  In this case our trial action $S_0$ consists of
ultra local terms. The partition function then factorizes into a
product of $N$ integrals over the fields at each lattice site.

Let us first consider the Free energy density, $F$.  Using
$\delta$ as a counting parameter and define
\begin{equation}\label{free}
F := -\frac{1}{N}\ln{Z}=-\frac{1}{N}\left
(\ln[Z_{0}]
+
\sum_{n=1}^{\infty} \frac{\delta^{n}}{n!} K_{n}
\right
)
\end{equation}
where we substituted equation \tref{delta} for $Z$.  It is straight
forward to equate powers of $\delta$ to get
a relation between the $K_n$'s,  the cumulant averages, and
the statistical averages of \tref{statav}, e.g.
\begin{equation}\label{K3}
K_3 = \ao{(S-S_0)^3} - 3 \ao{(S-S_0)^2} \ao{(S-S_0)} + 2
(\ao{(S-S_0)})^3 .
\end{equation}
The statistical averages required to calculate the $K_n$ can then
be seen to involve integrals over the fields at no more than
$(n+1)$ connected lattice sites.  Descriptions of the diagrammatic
method used to find all the relevant integrals are given
elsewhere, for instance see \cite{Wo,EJR98a,EJR98b}.

For the free energy we choose the unphysical variational
parameters, $j_\phi,j_\psi,k_\pm$, such that they minimize the
free energy (maximise the entropy).  For the free energy
calculation only, it is also the procedure that the principle of
minimal sensitivity would demand. After fixing the parameters, $F$
depends only on the parameters $m^{2}$ and $\lambda$.

We also studied the expectation values $\av{\phi}, \av{\psi},
\av{\phi^{2}}$ and $\av{\psi^{2}}$. To do this we again use the
cumulant expansion, which for the expectation value of an operator
$O$ can be expressed as
\begin{eqnarray}\label{delexp}
\av{O}&=& \frac{\prod_{i=1}^{N}\int \textrm{d} \phi_{i} \textrm{d}
\psi_{i} O e^{-S}}{\prod_{i=1}^{N}\int \textrm{d} \phi_{i}
\textrm{d} \psi_{i} e^{-S}}
\\
&=&
\frac{\sum_{n=0}^{\infty} \ao{(S_{0}-S)^n O} \delta^n/n!}{\sum_{n=0}^{\infty}  \ao{(S_{0}-S)^n}
\delta^n/n!}
=\sum_{n=1}^{\infty}\frac{\delta^{n-1}}{(n-1)!}L_{n}
\end{eqnarray}
Equating powers of $\delta$ produces the same type of relation
between statistical averages, involving simple integrals over $n$ connected
lattice sites, and the $L_n$'s, similar to that for $K_3$ in
\tref{K3}.
In practice we expand up to third order, $L_3$ and $K_3$. $\av{O}$ is
then a function of the physical parameters $m^{2},\lambda$ and the
variational ones, $j_{\phi},j_{\psi},k_+,k_-$,  but it does
\emph{not} depend on $N$ - the number of lattice sites.

At this point note that there are alternatives for the
optimization step used to set these variational parameters in
these expectation values.  The principle of minimal sensitivity,
used with great success in many circumstances and in particular in
\cite{Jo,AJ93a,AJ93b,AJP}, is based on \tref{PMS} where $\alpha
\in \{ j_{\phi},j_{\psi},k_+,k_- \}$.  We applied this method to
$\av{\phi}, \av{\psi}, \av{\phi^{2}}$ and $\av{\psi^{2}}$.
Anticipating the role that the symmetry plays (see discussion
below) we also studied the O(2) invariants $\av{\phi}^2 +
\av{\psi}^2$ and $\av{\phi^2} + \av{\psi^2}$.  However we were
unable to find any suitable solutions for the variational
parameters when we tried to minimise the variation in the
expectation values of the field with respect to the variational
parameters.  Therefore we reverted to the method used in
\cite{WZZSDYX}, and in calculations of all expectation values we
choose unphysical variational parameters, $j_\phi,j_\psi,k_\pm$,
which minimized the {\em free energy} for the same set of physical
parameters $m^{2},\lambda$.

The resulting expressions for each quantity involve a sum of about
100 simple terms, each made up of a product of various variational
and physical parameters and a few of a family of twenty six
different integrals.  A typical term contribution to the sum which makes
$K_3$ in $d$ space-time dimensions is\tnote{This is a term in a142 of the
free energy programme.}
\begin{equation}
12 d \, j_{\psi} \half (m^2 + k_+ - k_- ) I_{10} I_{13}
(I_{00})^{-2}
\end{equation}
where
\begin{eqnarray}
I_{ab} &=& \int_{-\infty}^{\infty} dx dy \; x^a y^b
\exp \{ j_{\phi} x + \half (k_+ + k_-) x^2 + j_\psi y + \half (k_+ - k_-)y^2 - \lambda x^2 y^2
\}.
\end{eqnarray}
All the integrals are of this form and so they are easily
calculated numerically.  The computational time consuming part was
finding the location of the minimum of the free energy or of an
expectation value to fix the variational parameters. This was done
numerically in the four-dimensional space of unphysical
parameters.  Though no numerical algorithm can guarantee that it
has found the global minimum, in practice it turned out that our
expressions were amenable to the simplest algorithm.  Overall the
computations took no more than a couple of days on a personal
computer.

It is important to check our results as the expressions are simple
but long, so errors can easily creep in. One check we used was to
set in turn each of the fields to zero. The model then reduces to
that of a single real field with a $\phi^4$ interaction and we
find that we reproduce the results of Wu et al.\ \cite{WZZSDYX}
who used the same approach and an $S_0$ equivalent in this limit.
They in turn showed that these results match lattice Monte Carlo
calculations extremely well. Another check, as we shall discuss
below, is to restrict the analysis to a subset of one or two
variational parameters.  If we do that then the results agree with
the analysis in the full four-dimensional $j,k$ space  within
numerical errors.\tnote{Well perhaps the errors are larger than
the expected numerical errors but the agreement is still very
good.}\tnote{Yet another check is that the whole approach is
extremely sensitive to small errors in the formulae since the
variational parameters are set through a highly non-linear
self-consistent procedure.  A couple of such errors in the code
were trapped this way.  }

\begin{figure}[htbp]
\setlength{\unitlength}{1in}
\begin{picture}(6,6)
\put(0,3){\includegraphics{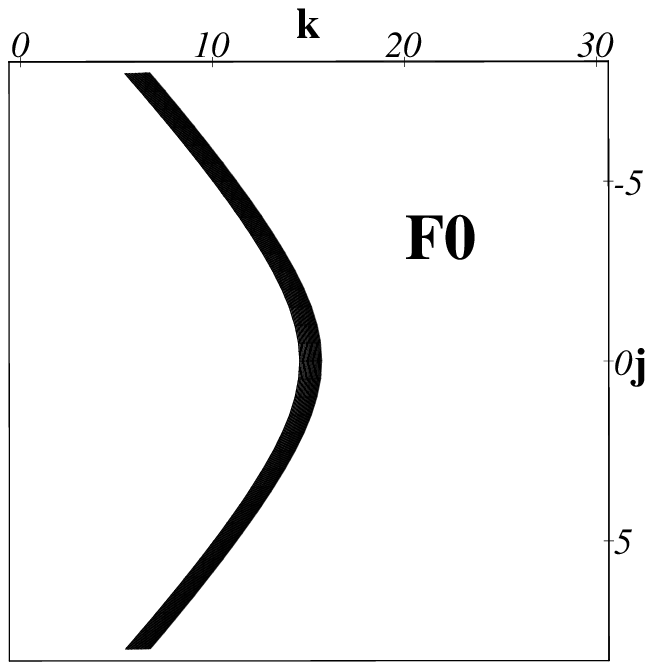}}
\put(3,3){\includegraphics{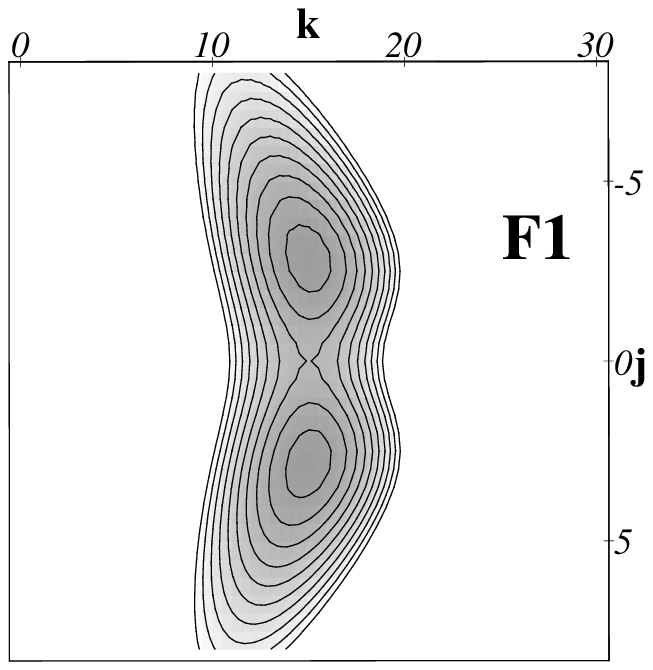}}
\put(0,0){\includegraphics{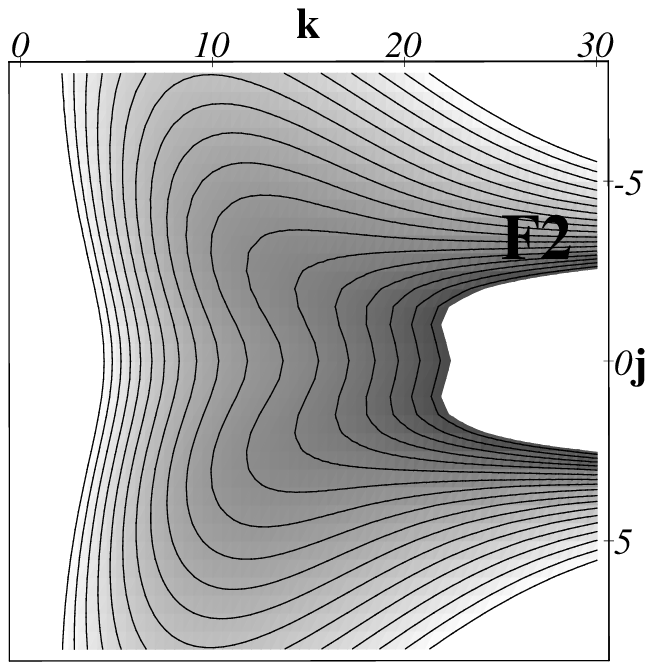}}
\put(3,0){\includegraphics{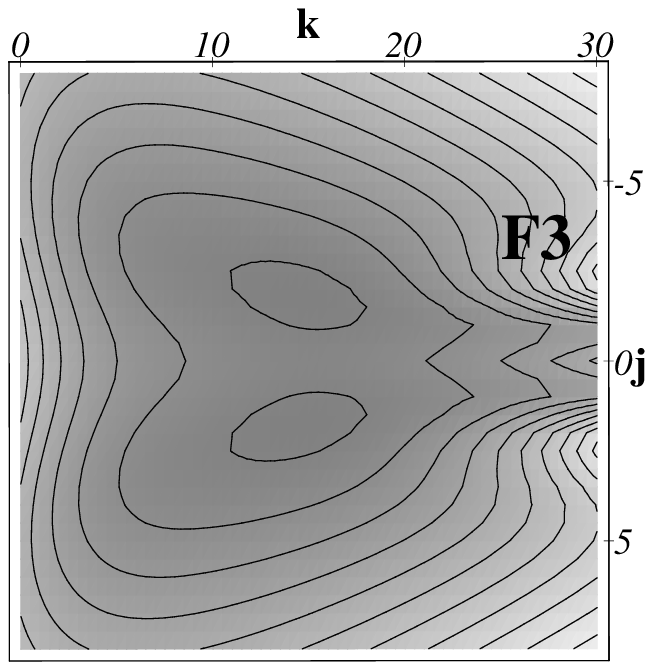}}
\end{picture}
\caption{Plots of the Free Energy from order zero (F0) to three
(F3) against $j=j_\phi$ and $k=k_+$ for $\lambda=25$, $m^2=-15$,
$j_\psi=0$ and $k_-=0$.  Contours are evenly spaced in the
interval -2.2 to -2.55, and show that only the odd orders have
minima which get shallower at higher orders.} \label{fe}
\end{figure}

\begin{figure}[htbp]
\setlength{\unitlength}{1in}
\begin{center}
\begin{picture}(3,3)
\put(0,0){\includegraphics{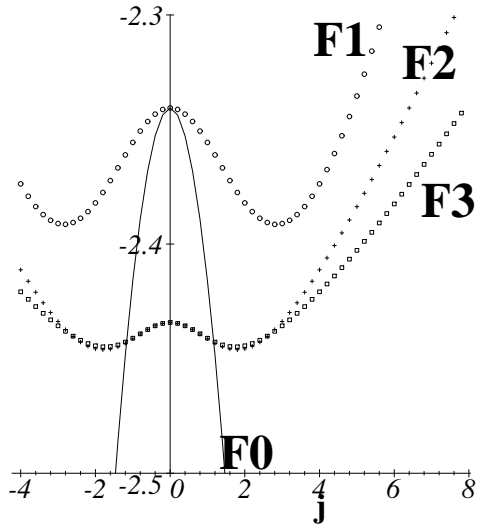}}
\end{picture}
\end{center}
\caption{Plots of the Free Energy from order zero (F0) to three
(F3) against $j=j_\phi$ for $\lambda=25$, $k_+=-m^2=15$,
$j_\psi=k_-=0$.}
\label{fefixk}
\end{figure}

The results for $\lambda=25$ are summarized in the figures. The
numerical errors are too small to see in the plots but the error
inherent in the method can be estimated by the difference between
the results for the first and third orders. First let us look at
the results for the free energy as a function of the variational
parameters.  It is sufficient to study the free energy with
$j_\psi=0$ and $k_-=0$ as we will show below, so in figure
\ref{fe} we show results for varying $j_\phi$ and $k_+$ for
$\lambda=25.0$ and $m^2=-15.0$.
First note that we find no minima or any turning points in the
even orders.  However, figure \ref{fe} shows that this is due
solely to the behaviour of the $k_+$ variational parameter which
is not always included in previous studies.\tnote{Which ones do
not have this?}.  In the $|j|$ variable alone, one does seem to find
clear minima as shown in figure \ref{fefixk}.
On the other hand, in the full variational parameter space there
are clear minima (and not just turning points demanded by the PMS
condition \tref{PMS}) in the first and third order, e.g.\ for
$\lambda=25$, $m^2=-15.0$ these are at $k_+ \approx -0.98m^2$ and
$|j| \approx 1.77$.\tnote{Taken from {\tt info13.txt} 22-03-99.}   This
difference in behaviour of even and odd orders is common in OPE
methods e.g.\ appears in \cite{EJR98a,EJR98b}. Further, the minima
in the first and third orders do not move much from the first to
third order but the minima get shallower at the higher order,
i.e.\ the results become less sensitive to the precise value used
for the variational parameters.  This is characteristic of good
PMS points.

Let us briefly consider the principle of fastest convergence of
\cite{KM} in this context.  This suggests that one should fix the
variational parameters by minimising the difference between
different higher and lower order terms in the $\delta$-expansion
in \tref{free}.  It is immediately obvious that this principle is
of no use when we have a function of more than one variational
parameter, as it provides only too few equations to constrain the
several variational parameters. However, we can arbitrarily reduce
the problem to a one-dimensional one by removing the $k$
variational parameters, i.e.\ set $k_+=-m^2$ and $k_-=0$ so that
no change in the quadratic part of the action is allowed.  These
values are also close to the values actually found to minimise the
free energy.  We then set $j_\psi=0$ by symmetry (see below)
leaving just one free parameter, the source $j_\phi$.  We are then
working with an $S_0$ similar to that used in some of the
literature.\tnote{Which ones?}  The results are plotted in figure
\ref{fefixk} for example values of $\lambda=25$ and $m^2=-15$
where we are in the spontaneous symmetry breaking regime. They
show that even in this one-dimensional subspace, the principle of
fastest convergence is still not working well for this Higgs
model.  The first and third order curves do not intersect but the
second and third orders do, contradicting normal linear
$\delta$-expansion behaviour where one compares every second
order. However the second and third orders do intersect but at a
value of $j_\phi$ much bigger than the location of the minimum. In
any case the difference between the two orders changes much more
slowly around the $j=0$ intersection\footnote{Diagramatically,
since there are no odd vertices when $j=0$, odd orders give zero
contributions at $j=0$.} suggesting that $j=0$ should be chosen by
this criterion. It is therefore difficult to see any reasonable
way in which the principle of fastest convergence of \cite{KM} can
be used to fix the variational parameters.

We do note that the plot in figure \ref{fefixk} does show that the
difference between the the second and third orders is still small
in the region where the Free energy is minimised (which is near
$k_+=-m^2$ and $k_-=0$). Thus using the minimum free energy
principle does give a series with higher order terms giving
successively smaller contributions, in the spirit of the principle
of fastest convergence.

\begin{figure}[htbp]
\begin{center}
\includegraphics{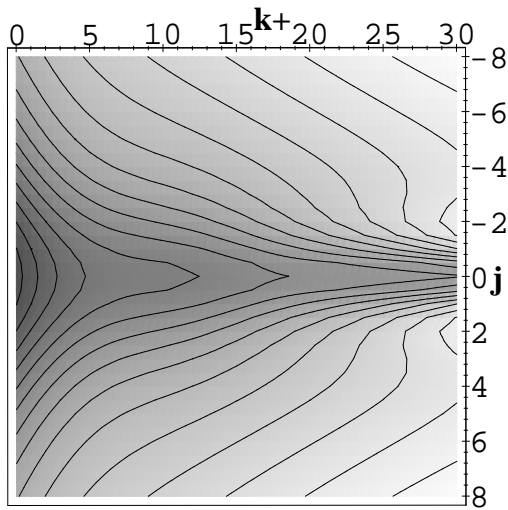}
\end{center}
\caption{Contour plot of $\av{\phi^{2}}+\av{\psi^{2}}$
against $j=j_\phi$ and $k_+$ at $\lambda=25$, $m^2=-15$, $j_\psi=k_-=0$.} \label{exjk}
\end{figure}

Now let us analyse the expectation values. One can just use the
values for the variational parameters which minimise the Free
energy, i.e.\ call upon some sort of maximum entropy principle. On
the other hand the very successful PMS approach \cite{St} would
demand that we look for turning points in the expectation values
with respect to the variational parameters \tref{PMS} and use
these values for the particular expectation value needed.  While
that was successful in gauged models
\cite{Jo,AJ93a,AJ93b,AJP,EJR98a,EJR98b}, we have not been able to
find any reasonable turning points in the expectation values of
the fields or fields squared, or even to O(2) invariant
combinations of expectation values.\tnote{Is this due to the $k+$
behaviour again?}  For instance, see figure
\ref{exjk}.\footnote{There are some turning points but they occur
in regions where the values are changing rapidly.  There was one
shallow saddle point in the $\av{\phi}^{2}+\av{\psi}^{2}$ plot at
$\lambda=25.0$ and $m^2=-15$, near $k_-=25$, $|j|=2$.  However no
other expectation value at any order showed any good behaviour in
this region and this suggests that this was a feature of
increasing ripples in the function as $k_+$ increased.}

Thus a major result of this paper is the realisation that the
previously successful PMS condition does {\em not} work for this pure Higgs
model.  However a minimum Free energy criterion does seem to give
well behaved results, i.e.\ weak dependence on the variational
parameters near the chosen solution, small corrections in the
variational parameters and free energy values when moving from
first to third orders.  There is no reason to believe that this
would not be the case in gauged models, at least for weak gauge
coupling.

\begin{figure}[htbp]
\begin{center}
\scalebox{0.5}{\includegraphics{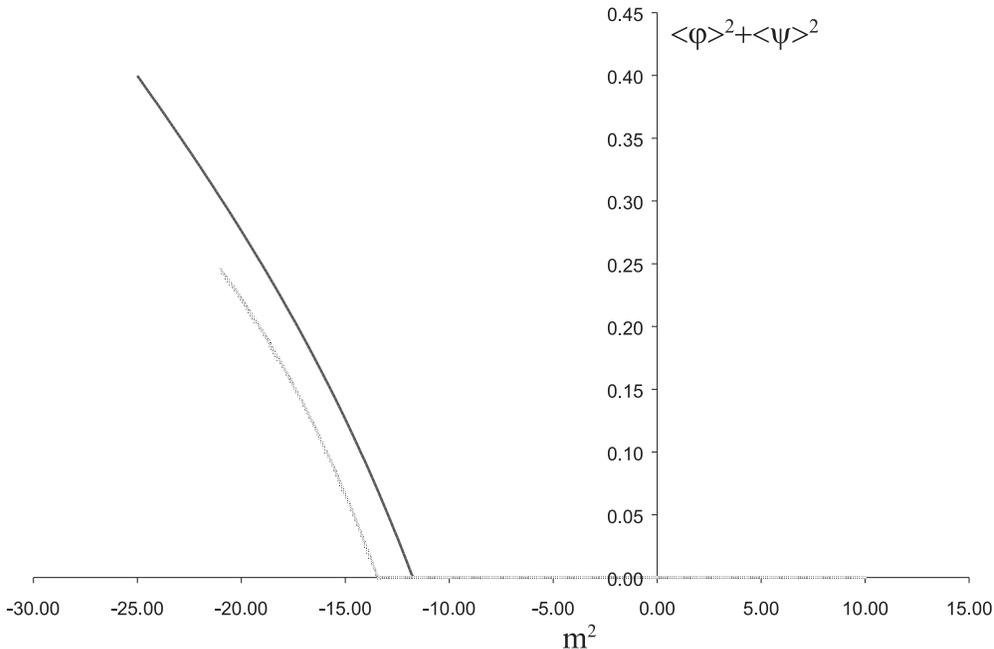}}
\caption{This plot shows the first and third order result of
$\av{\phi}^{2}+\av{\psi}^{2}$ vs. $m^{2}$ for $\lambda=25$. The
phase transition occurs around $m^{2} \approx -13.4$ for third
order and $m^{2} \approx -11.8$ for first order.} \label{phasetrans}
\end{center}
\end{figure}

\begin{figure}[htbp]
\begin{center}
\scalebox{0.5}{\includegraphics{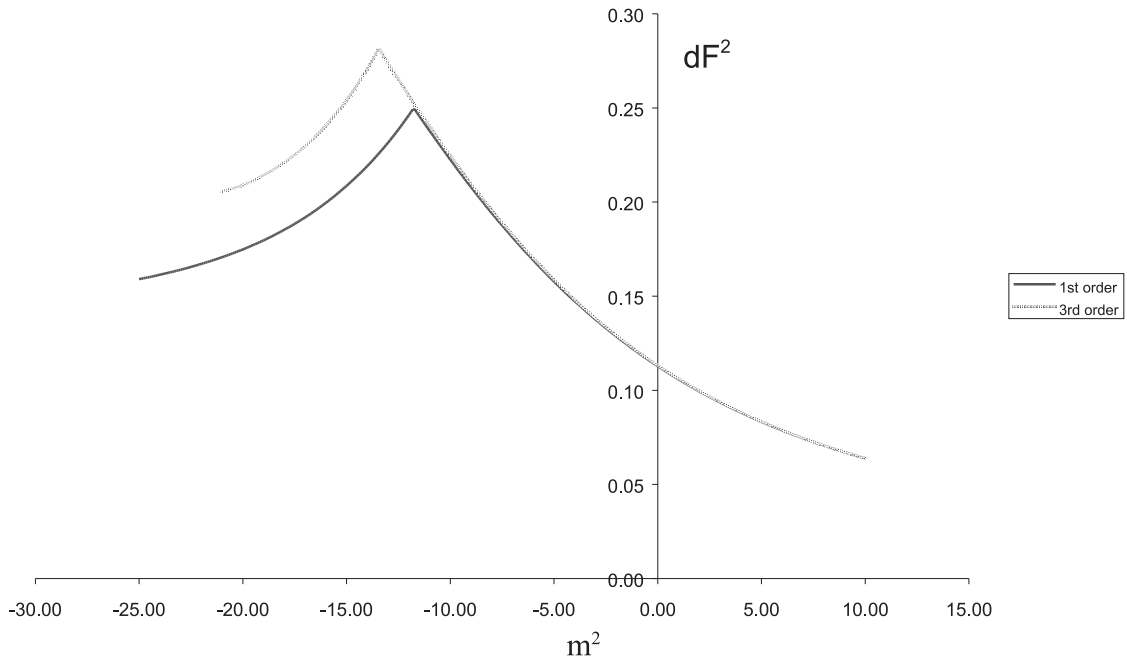}}
\caption{This plot shows the first and third order result of $dF^2
:=(\av{\phi^{2}}-\av{\phi}^{2}) + (\av{\psi^{2}} -\av{\psi}^{2})$
vs. $m^{2}$ for $\lambda=25$.} \label{fluct}
\end{center}
\end{figure}

Now let us try to use the variational values obtained from the
minimal Free energy principle to study the physics in this model
for the large coupling $\lambda=25$. The most prominent feature of the
complex $\phi^4$ theory is the phase transition. Figure
\ref{phasetrans} shows a plot of the quantity $\expectn {\phi}^2 +
\expectn {\psi}^2$ vs. $m^2$.
We can clearly see that the curves are continuous but that there
is a sudden change in slope at $m^2 \approx -13.4$ for the third
order expansion (and at $m^2 \approx -11.8$ for the first order
expansion).  This indicates a {\em second order} phase
transition.\tnote{A second order phase transition is one in which
the derivative is discontinuous (manifesting itself as a `kink' in
the function), while in a first order phase transition the
discontinuity appears in the function itself.} Since we are
working on a lattice and have not taken the continuum limit, the
mass parameter $m$ is not a physical mass so the fact that the
phase transition does not occur at $m^2=0$ (as it would
classically)\tnote{What about the continuum value for zero mass?}
is not surprising.  In any case we are working a long way from the
perturbative regime. The fluctuations about the minimum can be
estimated from the difference $(\expectn {\phi^2}- \expectn
{\phi}^2) + (\expectn {\psi^2}- \expectn {\psi}^2)$ and this is
shown in figure \ref{fluct}.\tnote{Are the comments which follow
OK?}  This peaks at the phase transition as expected.

Another point to note is the way that non-zero results for
$\expectn {\phi}$ and $\expectn {\psi}$ can be found in the OPE.
This is a key difference between this method and lattice Monte
Carlo methods. The OPE will always find one and only one vacuum
solution even if there is more than one solution.  It does this by
allowing the sources $j$ to take non-zero values in which case the
classical potential is tilted favouring one vacuum solution.
This contrasts with lattice Monte Carlo methods which sample all possible vacuum states with
equal likelihood. This means OPE methods are ideal for studying
non-trivial field configurations, perhaps with defects present, by
using classical sources to manipulate the effective quantum
potential.\tnote{Can we construct the effective potential?}

\begin{figure}[htbp]
\begin{center}
\scalebox{0.5}{\includegraphics{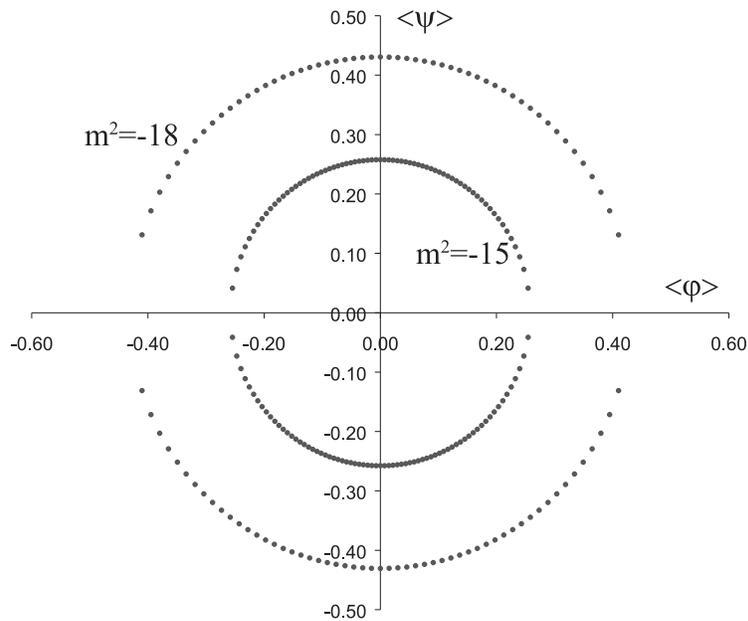}}
\caption{Plot of $\expectn {\psi}$ vs. $\expectn {\phi}$ for
$m^2=-15.0$ and $m^2=-18.0$, for various initial values for the
variational parameters. $\lambda=25.0$.
We see that there are many degenerate
vacuum states for a given $m^2$ in the broken symmetry regime.}
\label{circles}
\end{center}
\end{figure}

The last plot, figure \ref{circles}, clearly shows the role of the
O(2) symmetry in this method.
The plot of
$\expectn {\psi}$ vs. $\expectn {\phi}$ for $m^2=-15.0$ (inner curve)
and $m^2=-18.0$ (outer curve) both show circles, i.e.\ the
quantity $\expectn {\phi}^2 + \expectn {\psi}^2$ is a constant.

\begin{figure}[htbp]
\begin{center}
\scalebox{0.5}{\includegraphics{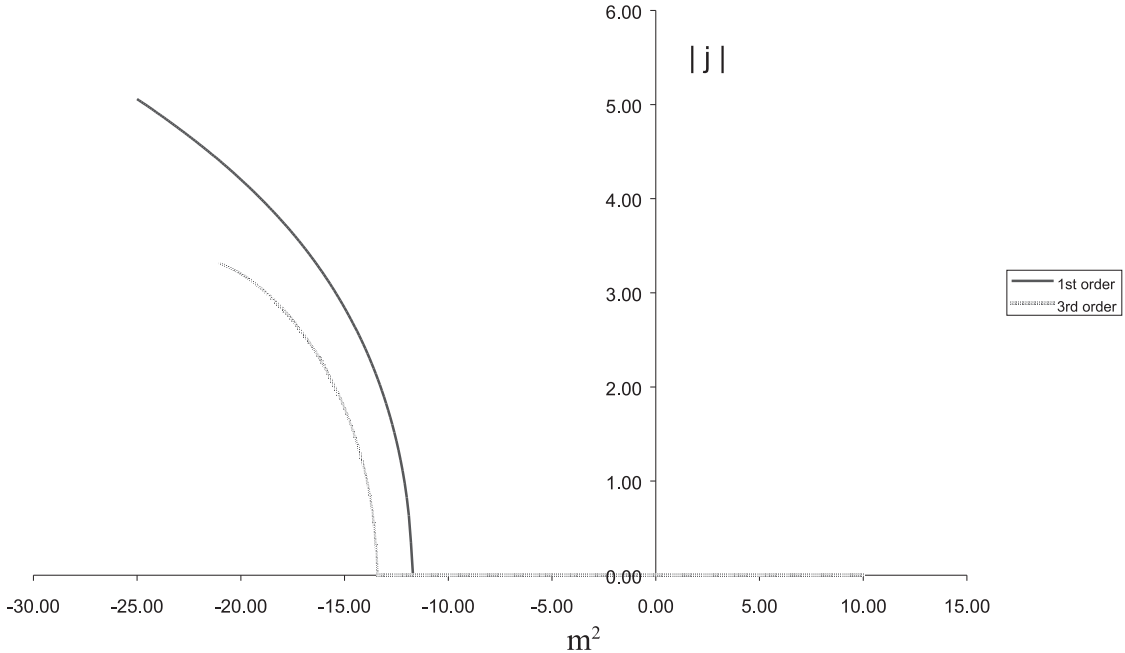}}
\caption{Plot of $|j|$ vs. $m^2$ for $\lambda=25.0$.
It shows that this variational parameter acts as an order parameter.}
\label{modj}
\end{center}
\end{figure}

It is important to note the form of the solutions found
for the unphysical parameters $j_\phi,j_\psi, k_+$ and $k_-$.  In
the unbroken phase, $j_\phi=j_\psi=0$.  In the broken phase while
$|j|^2 = j_\phi^2 + j_\psi^2$ is a non-zero constant for a given
set of physical parameters $m^2$ and $\lambda$, with a different
solution for the $j$'s corresponding to each point on the circles
in figure \ref{circles}.  This immediately suggests that $|j|$
acts as an order parameter.  Thus a calculation of just
the free energy, and by implication the variational parameters,
rather than of any expectation value can be used find the location
of the transition, as shown in figure \tref{modj}.

\begin{figure}[htbp]
\begin{center}
\includegraphics{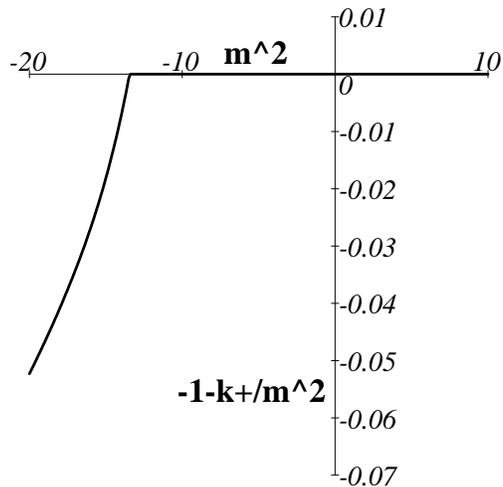}
\end{center}
\caption{Plot of $(-1-k_+/m^2)$ which minimise the third order free
energy vs.\ $m^2$ for $\lambda=25.0$.} \label{kpm2}
\end{figure}

The solution for $k_+$ was equal to $m^2$ in the unbroken regime
and within 10\% of $m^2$ otherwise, as shown in
figure \ref{kpm2}.
Thus $(K_+ + m^2)$ is acting as another order parameter.  This shows
that its numerical value, chosen by minimising the free energy, is
keeping $S$ and $S_0$ as similar to each other as possible. However,
it is not so much the numerical value of $k_+$ but the behaviour
of functions as we vary $k_+$ which is important.  As remarked
above, there is often no minimum in the $k_+$ variable when there
are minima in the other variational parameters.

Finally we note that that $k_-=0$
always minimised the free energy, within numerical accuracy.  This
latter result is of particular significance with regards the
symmetry of our ansatz $S_0$, and we will now turn to consider
this aspect.

To explain these results, in particular figure \ref{circles}, we
must consider the symmetry of the problem. $S$ is invariant under
O(2) transformations such as
\begin{equation}\label{phitrans}
\phivec \rightarrow \phivec' = U \phivec, \; \; \;
U=
\left( \begin{array}{cc} \cos ( \theta ) & \sin ( \theta ) \\
-\sin ( \theta ) & \cos ( \theta ) \end{array} \right) , \; \; \;
\phivec = \left( \begin{array}{c} \phi \\ \psi \end{array} \right) , \; \; \;
\end{equation}
Our $S_0$ of \tref{TrialS} is not O(2) invariant if $k_- \neq 0$
or if the $j$ parameters are non-zero.  However it is easy to see that
\begin{equation}\label{jtrans}
  S_0 (\phivec ; m^2, \lambda ; \jvec , k_+, k_-=0 ) =
  S_0 (\phivec' = U \phivec ; m^2, \lambda ; \jvec'=U \jvec , k_+, k_-=0 )
\end{equation}
where $\jvec=(j_\phi,j_\psi)$.  That is, there is a set of field
independent transformations for $j_\phi,j_\psi$ and $k_+$ which
leave $S_0$ invariant under the O(2) field transformation.  Note
that we {\em must} have $k_-=0$ for this to be true, so let us for
the moment assume that this value minimises the free
energy\footnote{Remember in our choice of $S_0$ in \tref{TrialS}
we used the O(2) field rotation to eliminate a possible
$k\phi\psi$ term.  However, with $k_-=0$ we regain the ability to
exploit this freedom.}.

Using this O(2) transformation we note that the integration
measure is also O(2) invariant.  It is then easy to show that at
whatever order in $\delta$ we truncate our free energy expression,
if the set $\{ \tilde{\jvec},\tilde{k}_+,\tilde{k}_-=0\}$
minimises the free energy then the set $\{ \tilde{\jvec'},
\tilde{k}_+, \tilde{k}_-=0\}$ also minimises the free energy.  It
then follows that a calculation of any O(2) invariant physical
object will not depend on which solution for the variational
parameters we use. In particular the free energy, $\expectn
{\phi^2} + \expectn {\psi^2}$, and $\expectn {\phi}^2 + \expectn
{\psi}^2$ will all give results which are independent of the
solution for the variational parameters.  For objects like
$\expectn {\phi}$ which are not O(2) invariant, we can still
relate one solution found for it, with a particular set of
variational parameters, to another solution for $\expectn {\phi}$
and another set of variational parameters.  The relationship is
precisely the simple O(2) transformations discussed here.

Plotting out final values for the variational parameters
$j_{\phi}$ and $j_ {\psi}$ for the different initial conditions
used in producing figure \tref{circles} we found that  $j_ {\phi}$
and $j_{\psi}$ also form circles in a $j_ {\phi}$-$j_ {\psi}$
plane (for fixed $m^2$ and $\lambda$).\tnote{Is there a simple
relation between the numerical value of $j$'s and classical
solutions? c.f. the one for $k_+$?} Actually, there is a
one-to-one correspondence between pairs of $j_{\phi}$, $j_{\psi}$
and \expectn {\phi}, \expectn {\psi} \--- two points separated by
a certain angle on the $j$ circle correspond to two points
separated by the same angle on the expectation value circle; so to
get one of the points from the circles in figure \ref{circles}, we
fix $j_{\phi}$ to a certain value, and minimize $F$ in the
remaining $3$-dimensional parameter space. Once the point is
found, we fix $j_{\phi}$ to a different value, and minimize again,
thereby finding a different point.  This is the explicit display
that O(2) symmetry transforms both $j$'s and $\phi$'s with the
same rotation matrix. The fact that we found $k_+$ was constant
for all these solutions and that $k_-=0$ confirms the role of the
symmetry transformations \tref{phitrans} and \tref{jtrans}.

The presence of the variational parameter $k_-$, which does not
respect the O(2) symmetry even if we allow field independent
transformations, is a crucial distinction from the work of
\cite{EJR98a,EJR98b} and \cite{WZZSDYX}. It means that we have
allowed our model to break the bounds of the symmetry of $S$ if
the dynamics so choose.  Thus one of the most important numerical
results we have is that $k_-=0$ to numerical accuracy (at least
six significant figures).   Once this is known, only then do the
symmetry arguments of the previous paragraph explain the circles
of figure \ref{circles}.

This explicit display of the role of the symmetry in the solutions
suggests an interesting check, namely to \emph{impose} the
symmetry on the $S_0$. Thus, we ran the whole process again
holding $k_-=0$, i.e.\ minimizing in a $3$-dimensional parameter
space. We recovered the circle of $j$'s (at fixed
$m^2$).\tnote{This time with a radius constant to 5 decimal
places, which is expected for a minimization done to 6 decimal
places. The radius of $\expectn {\phi}^2 + \expectn {\psi}^2$
increased its accuracy as well (from 5 to 6 decimal places) once
$k_-$ was set to zero.}  We can then take it a stage further as in
addition to holding the $k_-=0$, we can also fix one of the
$j$'s to zero, and thus minimize in a $2$-dimensional parameter
space. In the un-broken symmetry regime, the $j$'s are zero
anyway, while in the broken symmetry phase there is an infinite
set of equally valid solutions for the variational parameters, so
holding one of the $j$'s at zero does not imply any loss of
generality of the results.\tnote{Setting the sources to zero makes
a considerable simplification to the diagrammatic calculation.}

In principle, one of the big advantages of the OPE is that one can
systematically study higher orders through a straight forward
extension of the method used here. However, it is likely that no
solution for the variational parameters will be found at the next
order, just as none was found at second order.  As mentioned
earlier, this behaviour is common in an OPE.  Thus we would need
to calculate $K_5$ or $L_5$.\tnote{These are too complicated to do
by hand.  The approach would have to be automated which is
feasible by using some of the technology developed for Linked
Cluster Expansions, e.g.\ see T.Reitz.}.

To summarize, the OPE method used here works well in identifying
the phase transition when there is a continuous symmetry, provided
we fix the variational parameters by minimizing the free energy
and we do not use the previously successful principle of minimal
sensitivity. The explanation for the failure of the latter
approach in this case after its earlier successes seems to be our
use of a variational parameters which are quadratic in the Higgs
fields.\tnote{Did we include this term in \cite{EJR98a,EJR98b}?}
Another clear message coming from this work is that the optimal
solutions choose a trial action $S_0$ which has the same symmetry
as the full action $S$. This fact can be exploited to dramatically
reduce the number of variational parameters used, which in turn
will simplify and accelerate the analysis of more complicated
models.  Next, the source variational parameters, $j$, and the
combination $(k_++m^2)$, act as order parameters. Thus one can
identify the transition point from free energy calculations alone
which is a further simplification when trying to find the phase
diagram. We also note that this method can be used to study just
one of many equivalent vacuum solutions which distinguishes it
from lattice Monte Carlo. On the other hand, as shown here with
just three orders, the OPE has a practical and systematic way of
improving its accuracy, unlike other analytic non-perturbative
methods.  Thus we have demonstrated that OPE offers a
practical route to the study of models with Higgs sectors and
continuous symmetries in non-perturbative problems.

% --- Acknowledgement ---
\section*{Acknowledgements}

We would like to thank H.F.Jones and D.Winder for useful discussions.

% ******************************************************************

\tpre{

\section*{Appendix}

\begin{figure}[htbp]
\begin{center}
\scalebox{0.5}{\includegraphics{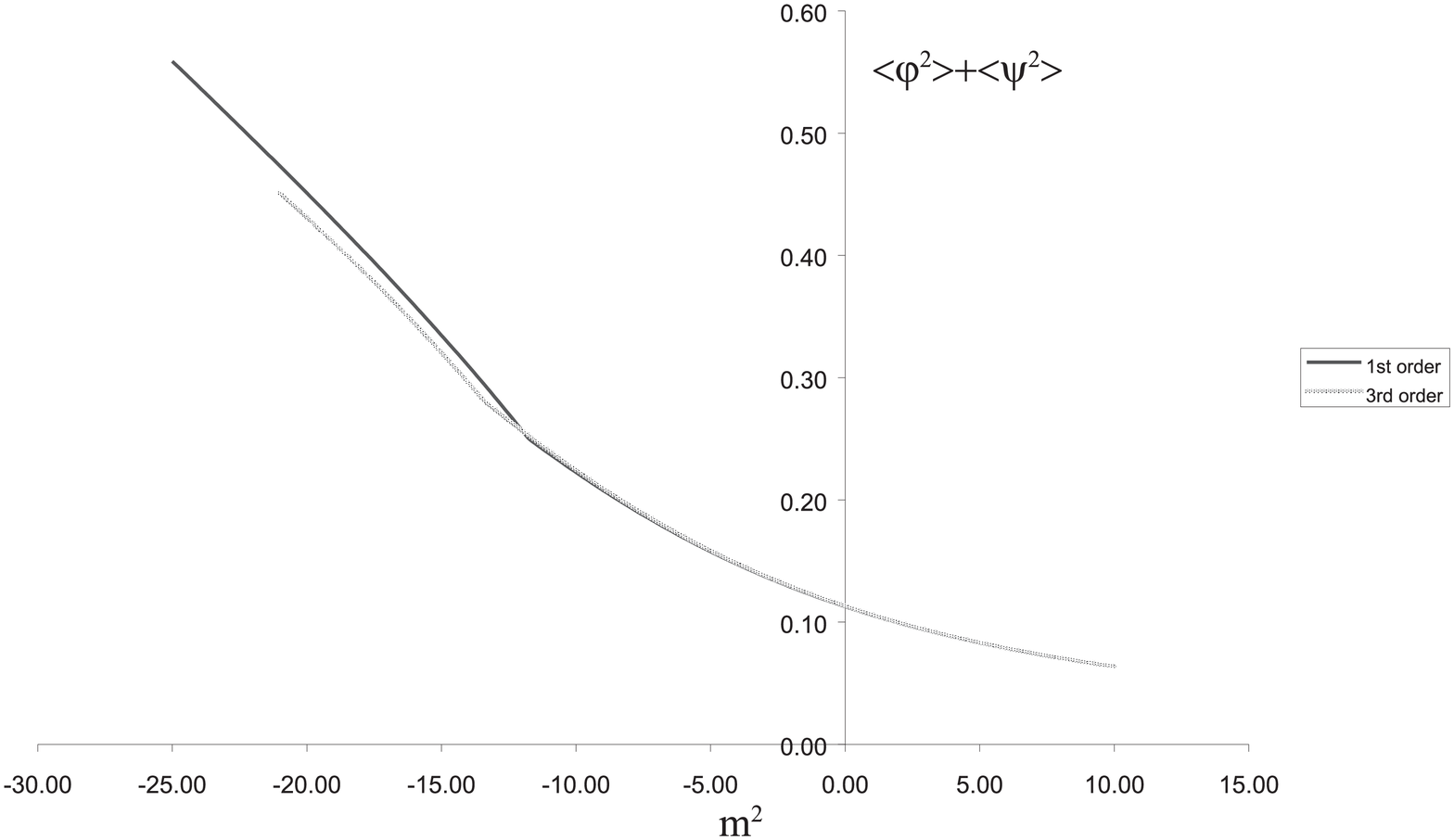}}
\caption{This plot shows the first and third order result of
$\av{\phi^{2}}+\av{\psi^{2}}$ vs. $m^{2}$ for $\lambda=25$. It
shows the phase transition, but the kink is not as clearly visible
as in other plots. } \label{phasetranssq}
\end{center}
\end{figure}
}

\end{document}